\documentstyle[12pt]{article}
\begin{document}

\begin{center}

{\sf \Large Nilpotent Symmetries and Curci-Ferrari Type Restrictions in 2$D$ Non-Abelian Gauge Theory: Superfield Approach}

\vskip 3.2 cm

{\sf{N. Srinivas$^{(a)}$, R. P. Malik$^{(a,b)}$}}\\
$^{(a)}$ {\it Physics Department, Institute of Science,}\\
{\it Banaras Hindu University, Varanasi - 221 005, (U.P.), India}\\

\vskip 0.1cm


$^{(b)}$ {\it DST Centre for Interdisciplinary Mathematical Sciences,}\\
{\it Institute of Science, Banaras Hindu University, Varanasi - 221 005, India}\\
{\small {\sf {E-mails:seenunamani@gmail.com; rpmalik1995@gmail.com}}}
\end{center}

\vskip 3.0cm

\noindent
{\sf Abstract:} We derive the off-shell nilpotent symmetries of the two (1+1)-dimensional (2{\it D})
non-Abelian 1-form gauge theory by using the theoretical techniques of the
geometrical superfield approach to Becchi-Rouet-Stora-Tyutin (BRST) formalism. For this purpose, we exploit the augmented version of superfield approach
(AVSA) and derive 
theoretically useful nilpotent (anti-)BRST, (anti-)co-BRST symmetries and Curci-Ferrari  (CF)
type restrictions for the self-interacting 2{\it D} non-Abelian 1-form gauge theory
(where there is {\it no} interaction with matter fields). The derivation of the (anti-)co-BRST symmetries 
and {\it all} possible CF-type restrictions are completely {\it novel} results within the framework of
AVSA to BRST formalism where the {\it ordinary} 2{\it D} non-Abelian theory is generalized onto an
appropriately chosen (2, 2)-dimensional supermanifold.
The latter is parameterized by the superspace coordinates $Z^{M} = (x^{\mu}, \theta, \bar\theta)$
where $x^{\mu }$ (with $\mu = 0,1$) are the bosonic coordinates and a pair of Grassmannian variables 
$(\theta, \bar\theta)$ obey the relationships: 
$\theta^{2} = \bar\theta^{2} = 0,\,\, \theta\bar\theta + \bar\theta\theta = 0$.
\vskip 1.2cm


\noindent
{\it Keywords}: 2{\it D} non-Abelian gauge theory; augmented version of superfield approach; BRST formalism; 
(anti-)BRST and (anti-)co-BRST symmetries; 
Curci-Ferrari type restrictions

\newpage
\section{Introduction}

The principles of {\it local} gauge symmetries (and their consequences) are at the 
heart of the precise theoretical description
of three (out of four) fundamental interactions of nature [1]. The gauge theories, based
on the {\it above} local symmetries, are quantized covariantly and consistently within the framework
of Becchi-Rouet-Stora-Tyutin (BRST) formalism where the local gauge symmetries of the original {\it classical}
gauge theories are traded with the quantum gauge [i.e. (anti-)BRST] symmetries (at the {\it quantum} level). 
For a given local
gauge symmetry, there exist {\it two} quantum gauge symmetries (within the framework of BRST formalism) 
which are christened as the BRST
and anti-BRST symmetries. The nilpotency and absolute anticommutativity properties are the {\it two} 
decisive features of these symmetries which encompass in their folds the properties of ``supersymmetry'' and 
linear independence, respectively. In other words, the transformations generated by the (anti-) BRST
symmetries are fermionic (i.e. supersymmetric-type) in nature and they have their own independent
identities due to their absolute anticommutativity property.

Just like the key features of supersymmetric transformations, the (anti-)BRST symmetries transform 
a bosonic field into  fermionic field and {\it vice-versa}. Both types of symmetry transformations are
nilpotent of order two. 
There is a distinct difference between the
above cited two types of symmetry transformations, however. Whereas the BRST and anti-BRST symmetry transformations
(corresponding to a given local gauge symmetry) must anticommute, the two distinct supersymmetric transformations
do {\it not} absolutely anticommute. Rather, the anticommutator of latter two distinct transformations {\it always}
generates a spacetime translation of the field on which they operate. Thus, it is crystal clear that
the (anti-)BRST symmetry transformations (in the context of quantization of gauge theories) is {\it not} exactly
like the supersymmetric transformations despite the fact that both types of transformations are fermionic (i.e.
nilpotent of order two) in nature.

The usual superfield approach (USFA) to BRST  formalism [2-9] sheds light on the properties
of nilpotency and absolute anticommutativity because the (anti-)BRST symmetries are identified with
the {\it translational generators} along a pair of Grassmannain varibles ($\theta, \bar \theta$) which characterize
the ($D$, 2)-dimensional supermanifold on which a given $D$-dimensional 
{\it ordinary} gauge theory is generalized. To be precise, the ($D$, 2)-dimensional 
supermanifold is parametrized by the superspace variables 
$Z^{M} = (x^{\mu}, \theta, \bar\theta)$ where the bosonic coordinates $x^\mu$ (with $\mu = 0, 1, 2...D-1$)
correspond to the ordinary $D$-dimensional spacetime variables and the Grassmannian 
variables ($\theta, \bar\theta $) satisfy the standard 
relationships: $\theta^2 = \bar\theta^2 = 0, \theta \bar \theta + \bar\theta \theta = 0  $. 
In the above {\it identification} (and geometrical {\it interpretation}), the celebrated horizontality 
condition (HC) plays a key role which primarily leads to the derivation of (anti-)BRST 
transformations for the {\it gauge} fields and the corresponding {\it (anti-)ghost} 
fields of a given $D$-dimensional gauge theory (described within the framework of 
USFA to BRST formalism) and Curci-Ferrari type of restrictions.

The above USFA has been systematically and consistently generalized so as to derive the (anti-)BRST symmetry
transformations for the gauge, (anti-)ghost and {\it matter} fields {\it together} for an {\it interacting}
gauge theory where there is a coupling between the gauge fields and matter fields [10-14]. 
In the above generalization, in addition to the HC, we invoke additional 
restrictions [i.e. gauge invariant restriction (GIRs)] which
are consistent with the HC and there is an inter-relationship and inter-dependence between the HC and GIRs
in such a manner that the geometrical {\it interpretation} of the (anti-)BRST symmetries 
(and corresponding conserved charges)
remains intact. The generalized version of the superfield approach has been christened as the augmented version of superfield
approach (AVSA) to BRST formalism [10-14]. We have exploited the latter superfield approach (i.e. AVSA) to discuss the 
central theme of our present endeavor where we have derived the off-shell
nilpotent and absolutely anticommuting (anti-)BRST as well as (anti-)co-BRST
symmetries and {\it all} the possible Curci-Ferrari (CF) type restrictions that would, in general,
be supported by the 2$D$ non-Abelian gauge theory under consideration. 
In our present work, we have, however, utilized {\it only} a few of the {\it total} CF-type restrictions 
(supported by our 2$D$ theory).

In our present investigation, we concisely mention the results of [5,6] where we discuss
the strength of HC in the derivation of proper (anti-)BRST symmetries for the non-Abelian theory
(in any arbitrary dimension of spacetime). The novelty of our present work begins with the derivation of
proper (anti-)co-BRST
symmetry transformations where we exploit the virtues of AVSA to BRST formalism. In fact, we utilize
the ideas of dual-HC (DHC) and dual-gauge invariant restrictions (DGIRs) for the {\it complete} derivation of proper (anti-) co-BRST
symmetry transformations for {\it all} the fields of our theory. The highlights of our present investigation
are, however, the derivation of CF-type restrictions using the AVSA to BRST formalism where the inputs from
the results, obtained from the application of HC, DHC, GIRs, as well as DGIRs, are utilized {\it together}. 
We have been able to compute {\it all} possible CF-type restrictions from the original CF-condition 
$[B+\bar B+(C\times \bar C) = 0]$ by requiring the (anti-)BRST and (anti-)co-BRST invariance of it 
within the framework of geometrically rich AVSA to BRST formalism.

Our present investigation is inspired and influenced by the following key factors. First, the 2$D$ 
non-Abelian 1-form gauge theory (without any interaction with matter fields) is the only
{\it non-Abelian} 1-form gauge model where we have been able to demonstrate the existence of (anti-)dual BRST
[i.e. (anti-)co-BRST] symmetry transformations. Thus, it is challenging for us to derive these (anti-)co-BRST symmetry 
transformations from the AVSA. Second, the insights and understanding gained from our present 
endeavor would be useful in obtaining the (anti-)co-BRST symmetry transformations
for the higher $p$-form ($p=2,3....)$ gauge theories within the framework of AVSA. In this connection,
we mention that, for the 4$D$ Abelian 2-form and 6$D$
Abelian 3-form gauge theories, we have already shown the existence of the (anti-)BRST and (anti-)co-BRST
symmetry transformations {\it together} [15-17]. Finally, one of the key signatures of the BRST approach to
the $p$-form ($p=1,2,3,...$) gauge theories is the existence of the CF-type restrictions. Thus, it 
is a challenging problem for us to derive them within the framework of  AVSA (particularly in the cases where
the (anti-)BRST and (anti-)co-BRST symmetries exist {\it together}). We have derived {\it all} possible
CF-type restrictions that could be supported by the 2$D$  non-Abelain theory where the (anti-)BRST and (anti-)co-BRST
symmetries {\it co-exist}. However, only {\it a few} of these have been actually used by us in
the discussion of the symmetries of our 2$D$ theory within the framework of BRST formalism.

We would like to comment on the existence of the (anti-)co-BRST symmetries
in the context of BRST approach to gauge theories. For the one (0+1)-dimensional
(1$D$) toy model of a rigid rotor, we have demonstrated that the (anti-)co-BRST
symmetries exist under which the gauge-fixing term remains invariant [15]. This
observation should be contrasted with the existence of the (anti-)BRST symmetries
under which the kinetic term remains invariant. We have established that the
nilpotent (anti-)co-BRST symmetries also exist for any arbitary Abelian $p$-form
($p = 1, 2, 3...$) gauge theory in $D = 2p$ dimensions of spacetime (see, e.g. [16]
and references therein). The decisive features of the (anti-)BRST and (anti-)co-BRST symmetries have 
been shown, once again, in the invariance of the kinetic
and gauge-fixing terms, respectively, for the above Abelian p-form gauge theories.
The existence of the nilpotent (anti-)co-BRST symmetries is physically important
because these have led to the proof that the 2$D$ (non-)Abelian 1-form gauge theories
belong to a new class of topological field theory (see, e.g. [21] and references therein
for details) and 4$D$ Abelain 2-form as well as 6$D$ Abelain 3-form gauge theories are
models of quasi-topological field theories (see, e.g. [16] for details).

The material of our present investigation is organized as follows. In Sec. 2, we discuss very 
concisely the (anti-)BRST
and (anti-)co-BRST symmetries for the 2{\it D} non-Abelian 1-form gauge theory in the Lagrangian formulation
to set-up the notations and convention. Our Sec. 3 is devoted to a brief synopsis of HC so that
our paper could be self-contained. The subject matter of Sec. 4 is the application 
of the AVSA to derive the (anti-)co-BRST symmetry transformations using the DHC
and DGIR for the 2{\it D} non-Abelian 1-form gauge
theory. Our Sec. 5 deals with the derivation of {\it all} possible CF-type
restrictions that could be supported by our 2$D$ theory by using AVSA to BRST formalism. Finally, we make some concluding remarks
in Sec. 6 and point out a few future directions for further investigations.

In our Appendix A, we discuss some explicit computations which have been incorporated 
in the main body of the text of our present endeavor.\\

\noindent  
{\it Notations and Convention}: Throughout the whole body of our text, we use the notations
$s_{(a)b}$ and $s_{(a)d}$ for the (anti-)BRST and (anti-)dual-BRST symmetry transformations. The covariant
derivative $D_{\mu}C = \partial_{\mu} C + i \,(A_{\mu}\times C)$ is in the adjoint representation of 
the {\it SU(N)} Lie algebraic space where the generators $T^{a}$  (with $ a = 1,2,...N^{2}-1$) obey the Lie algebra 
$[T^{a}, T^{b}] = f^{abc} \,T^{c}$. The structure constants $f^{abc}$ are chosen to be totally antisymmetric
for the semi-simple Lie group {\it SU(N)}. We further adopt the notations
$P\cdot Q = P^{a}\,Q^{a}$ and $(P\times Q)^{a} = f^{abc}\,P^{b}\,Q^{c}$ where the Latin indices 
$a,b,c....= 1,2,3...N^{2}-1 $ and $(P^{a}, Q^{a})$ are chosen to be non-null vectors in the Lie-algebraic space. We choose
the background 2{\it D} flat metric $\eta_{\mu\nu}$ with signatures (+1, -1) so that 
$A_{\mu}B^{\mu} = \eta_{\mu\nu}A^{\mu}B^{\nu} \equiv A_{0}B_{0} - A_{i}B_{i}$ where the Greek indices
$\mu, \nu, \lambda.... = 0, 1$ stand for the spacetime directions and the Latin indices
$i, j, k....=1$ correspond to the space direction {\it only}. In addition, the 2$D$ Levi-Civita tensor
$\varepsilon_{\mu\nu}$ has been chosen such that $\varepsilon_{01} = +1 = \varepsilon^{10}$ and
$\varepsilon_{\mu\nu}\varepsilon^{\mu\nu} = - 2!, \,\,
\varepsilon_{\mu\nu}\varepsilon^{\nu\lambda} = \delta_{\mu}^\lambda$, etc.

\section{Nilpotent Symmetries: Lagrangian Formulation}

Let us begin with the following coupled (but equivalent) (anti-)BRST invariant 
Lagrangian densities [18] for the 2{\it D} non-Abelian 
1-form $(A^{(1)} = dx^{\mu}A_{\mu}\cdot T) $ gauge theory in the Curci-Ferrari gauge (see, e.g [19, 20] for details)
\begin{eqnarray}
{\cal L}_B = - \frac{1}{4} F_{\mu\nu}\cdot F^{\mu\nu} + B \cdot (\partial_{\mu} A^{\mu}) 
+ \frac{1}{2}\,(B\cdot B + \bar B \cdot \bar B) - i \,\partial_{\mu}\bar C \cdot D^{\mu}C, \\ \nonumber
{\cal L}_{\bar B} = - \frac{1}{4} F_{\mu\nu}\cdot F^{\mu\nu} - {\bar B}\cdot (\partial_{\mu} A^{\mu}) 
+ \frac{1}{2}\,(B\cdot B + \bar B \cdot \bar B) - i\, D_{\mu}\bar C \cdot \partial^{\mu} C,  
\end{eqnarray}
where $B$ and $\bar B$ are the Nakanishi-Lautrup type auxiliary fields and anticommuting
$C^{a}\,\bar C^{a} + \bar C^{a} C^{a} = 0$ (anti-)ghost fields $(\bar C^{a}) C^{a}$ are fermionic
$[(C^{a})^{2} = (\bar C^{a})^{2} = 0]$ in nature. The 2-form 
$F^{(2)} = dA^{(1)} + i\, A^{(1)}\wedge A^{(1)} \equiv \bigl(dx^{\mu}\wedge dx^{\nu}/2!\bigr)\,(F_{\mu\nu}\cdot T)$
defines the curvature tensor $F_{\mu\nu} = \partial_\mu A_\nu - \partial_\nu A_\mu + i\,(A_\mu\times A_\nu)$ 
which has {\it only} one existing independent component in 2{\it D}, namely; 
$F_{01} = \partial_{0}A_{1} - \partial_{1}A_{0} + i\,(A_{0}\times A_{1}) \equiv E$.
Thus, for the case of 2{\it D} non-Abelian theory, we have the following coupled Lagrangian 
densities corresponding to (1), namely;  
\begin{eqnarray}
{\cal L}_B^{(2D)} =  \frac{1}{2} \,E \cdot E + B \cdot (\partial_{\mu} A^{\mu}) 
+ \frac{1}{2}\,(B\cdot B + \bar B \cdot \bar B) - i \,\partial_{\mu}\bar C \cdot D^{\mu}C, \\ \nonumber
{\cal L}_{\bar B}^{(2D)} = \frac{1}{2}\, E\cdot E - {\bar B}\cdot (\partial_{\mu} A^{\mu}) 
+ \frac{1}{2}\,(B\cdot B + \bar B \cdot \bar B) - i\, D_{\mu}\bar C \cdot \partial^{\mu} C. 
\end{eqnarray}  
The above Lagrangian densities are {\it equivalent} on the constrained hypersurface where
$ B\cdot (\partial_{\mu}A^{\mu}) - i\,\partial_{\mu}\bar C \cdot D^{\mu}C
= -B\cdot (\partial_{\mu}A^{\mu}) - i\, D_{\mu}\bar C \cdot \partial^{\mu} C$. This equality leads to the
CF-condition $ B + \bar B + (C\times \bar C) = 0$ (modulo a total spacetime derivative term).
The following supersymmetric-type (anti-)BRST transformations $(s_{(a)b})$  
\begin{eqnarray}
&&s_b A_\mu = D_\mu C, \qquad\qquad \;\;\;s_b C = -\frac{i}{2}\,\,(C\times C),\qquad \qquad s_b\bar C = i\, B, \\ \nonumber
&&s_b\bar B = i \,\,(\bar B \times C),\qquad \qquad 
s_b F_{\mu\nu} = i\,\, (F_{\mu\nu}\times C),\qquad \qquad s_b B = 0, \\ \nonumber
&&s_{ab} A_\mu = D_\mu \bar C,\quad \qquad s_{ab} \bar C = -\frac{i}{2}\,\,(\bar C\times \bar C), 
\qquad \qquad s_{ab} C = i\,\,\bar B, \\ \nonumber
&&s_{ab} B = i\,\, (B\times \bar C),\quad \qquad 
s_{ab} F_{\mu\nu} = i\,\,(F_{\mu\nu}\times \bar C),\qquad \qquad s_{ab} \bar B = 0, 
\end{eqnarray}  
are the {\it symmetry} transformations for the action integral 
$S = \int d^{2}x \,{\cal L}_B \equiv \int d^2x\,{\cal L}_{\bar B}  $ because one observes that the following are
true, namely;
\begin{eqnarray} 
s_b{\cal L}_B &= &\partial_\mu [B\cdot D^{\mu}C],\quad \qquad 
s_{ab}{\cal L}_{\bar B} = \partial_\mu [-\bar B\cdot D^{\mu}\bar C],
\nonumber \\
s_{ab}{\cal L}_B &=& \partial_\mu [- \{ \bar B + (C \times \bar C)\} \cdot \partial^{\mu}\bar C)] 
+ [B + \bar B + (C\times \bar C)]\cdot D_\mu\partial^{\mu}\bar C, \nonumber\\
s_{b}{\cal L}_{\bar B} &= &\partial_\mu [\{B + (C\times\bar C)\}\cdot\partial^{\mu}\bar C] 
- [B + \bar B + (C\times \bar C)]\cdot D_\mu\partial^{\mu} C. 
\end{eqnarray}  
These relationships establish that the above symmetries are {\it true} on a constrained hypersurface, 
embedded in the 2{\it D} spacetime
manifold, where the Curci-Ferrari (CF) condition $B + \bar B + (C\times \bar C) = 0$ is valid. It is 
elementary, at this stage, to note that we have $s_{ab}\,{\cal L}_B = \partial_\mu [B\cdot\partial^\mu\bar C]$
and $s_{b}\,{\cal L}_{\bar B} = -\, \partial_\mu [\bar B\cdot\partial^\mu C]$ due to the validity of CF-condition.
Further, we note that the absolute anticommutativity 
$(s_b s_{ab} + s_{ab} s_{b} = 0)$ property of the (anti-)BRST transformations $s_{(a)b}$ is true {\it only} 
when the (anti-)BRST invariant (i.e. $s_{(a)b}\,[B + \bar B + (C\times \bar C)] = 0$)
CF-condition $B + \bar B + (C\times \bar C) = 0$ is imposed from out side. It is crucial to point out that the 
gauge-fixing and Faddeev-Popov ghost terms of the starting Lagrangian density (1) can be written as [19,20]
\begin{eqnarray}
&&{\cal L}_{B} = - \frac{1}{4} \,F^{\mu\nu}\cdot F_{\mu\nu} 
+ s_{b}s_{ab} \Bigl (\frac{i}{2} \,A_{\mu}\cdot A_{\mu} - \frac{\xi }{2}\, \bar C\cdot C\Bigr ), \\ \nonumber
&&{\cal L}_{\bar B} = - \frac{1}{4} \,F^{\mu\nu}\cdot F_{\mu\nu} 
- s_{ab}s_{b} \Bigl (\frac{i}{2} \,A_{\mu}\cdot A_{\mu} - \frac{\xi }{2}\, \bar C\cdot C\Bigr ),
\end{eqnarray}
where the Curci-Ferrari gauge condition (cf. Eq. (1)) implies that we have chosen $\xi =2$.
Thus far, all our statements are true in {\it any} arbitrary dimension of spacetime. In other words,
the (anti-)BRST symmetries (3) are {\it true} for {\it any} arbitrary non-Abelain 1-form
gauge theory (when we discuss the theory within the framework of BRST formalism).

In addition to the above nilpotent and absolutely anticommuting (anti-)BRST symmetry 
transformations $s_{(a)b}$, we {\it also}
have a set of proper (i.e. nilpotent and absolutely anticommuting) (anti-)co-BRST symmetry
transformations $(s_{(a)d})$ in our theory. These transformations, in the context of our
2{\it D} non-Abelian theory, are (see, e.g. [21]):
\begin{eqnarray}
&&s_{ad}A_\mu = - \varepsilon_{\mu\nu}\partial^{\nu}C, \quad\qquad s_{ad}C = 0, \quad\qquad s_{ad}\bar C = i\,{\cal B},
\quad\qquad s_{ad}{\cal B} = 0, \\ \nonumber
&&s_{ad}E = D_{\mu}\partial^{\mu}C, \qquad s_{ad}(\partial_{\mu}A^{\mu}) = 0, \quad\qquad s_{ad}B = 0, 
\qquad s_{ad}\bar B = 0, \\ \nonumber
&&s_{d}A_{\mu} = -\varepsilon_{\mu\nu}\partial^{\nu}\bar C, \quad\qquad s_{d}\bar C = 0, \quad\qquad
s_{d}C = - i\,{\cal B}, \quad\qquad s_{d}{\cal B} = 0, \\ \nonumber
&&s_{d} E = D_{\mu}\partial^{\mu}\bar C, \qquad s_{d}(\partial_{\mu}A^{\mu}) = 0,\quad\qquad s_{d}B = 0,
\quad\qquad s_{d}\bar B = 0.
\end{eqnarray}  
The above off-shell nilpotent (anti-)co-BRST transformations are the {\it symmetry} transformations of the following
Lagrangian densities  
\begin{eqnarray}
&&{\cal L}_ B^{(2D)} = {\cal B} \cdot E - \frac{1}{2}\, {\cal B}\cdot {\cal B} + B\cdot (\partial_{\mu}A^{\mu})
+ \frac {1}{2}(B\cdot B + \bar B \cdot \bar B) - i \,\partial_{\mu}\bar C \cdot D_{\mu} C, \\ \nonumber
&&{\cal L}_{\bar B}^{(2D)} = {\cal B} \cdot E - \frac{1}{2}\, {\cal B}\cdot {\cal B} 
- \bar B\cdot (\partial_{\mu}A^{\mu})
+ \frac {1}{2}(B\cdot B + \bar B \cdot \bar B) - i\, D_{\mu}\bar C \cdot \partial^{\mu} C,
\end{eqnarray} 
where we have linearized the kinetic term $(\frac {1}{2} E \cdot E)$ of the Lagrangian density (2) by 
introducing the auxiliary field ${\cal B}$. It is straightforward to check that the (anti-)co-BRST
symmetry transfromations $(s_{(a)d})$ are off-shell nilpotent $(s_{(a)d}^{2} = 0)$ of order two and they
are absolutely anticommuting $(s_{d}s_{ad} + s_{ad}s_{d} = 0)$. The former property ensures the fermionic
(supersymmtric) nature of $s_{(a)d}$ and the latter property encodes the linear independence
of $s_{d}$ and $s_{ad}$. It can be readily checked that 
$s_{d} {\cal L}_{ B}^{(2 D)} = \partial_{\mu} [{\cal B}\cdot \partial^{\mu}\bar C]$ and 
$s_{ad} {{\cal L}_{\bar{  B}}}^{(2 D)} = \partial_{\mu} [{\cal B}\cdot \partial^{\mu} C]$. 
Hence the action integrals $S = \int d^{2}x \,{\cal L}_B^{(2D)} = \int d^{2}x \, {\cal L}_{\bar{ B}}^{(2D)}$
remain invariant under $s_{(a)d}$.

We close this section with the remark that we also end up with CF-type of restrictions (corresponding
to the (anti-)co-BRST symmetry transformations) when $s_{ad}$ and $s_d$ are applied on 
${\cal L}_B$ and ${\cal L}_{\bar B}$, respectively. In other words, we have the following [22]
\begin{eqnarray}
s_d{\cal L}_{\bar B}& =& \partial_{\mu}\bigl[{\cal B}\cdot D^{\mu}\bar C  
-\varepsilon^{\mu\nu}C\cdot\partial_\nu \bar C\times \bar C\bigr ] 
+ i\,\,(\partial_{\mu}A^\mu)\cdot ({\cal B}\times \bar C), \\ \nonumber 
s_{ad}{\cal L}_B &=& \partial_{\mu}\bigl[{\cal B}\cdot D^{\mu} C  
+\varepsilon^{\mu\nu}\bar C\cdot\partial_\nu  C\times  C\bigr ] 
+ i\,\,(\partial_{\mu}A^\mu)\cdot ({\cal B}\times  C), 
\end{eqnarray}
which lead to the existence of the following CF-type restrictions: 
\begin{eqnarray}
{\cal B}\times C = 0,  \quad\qquad {\cal B}\times \bar C = 0.
\end{eqnarray}
These restrictions are (anti-)co-BRST invariant
[i.e. $s_{(a)d}({\cal B}\times C) = 0$, $s_{(a)d}({\cal B}\times \bar C) = 0$].
Thus, we note that the CF-type restrictions (${\cal B}\times C = 0,\,\, {\cal B}\times \bar C = 0$),
in the context of (anti-)co-BRST symmetries, are {\it different} from the CF-condition
$B + \bar B + (C\times \bar C) = 0$ related with the (anti-)BRST symmetry transformations in
the sense that:
\begin{eqnarray}
&&s_{ab}\,[B + \bar B + (C\times \bar C) ] = i\,[B + \bar B + (C\times \bar C)]\times \bar C, \nonumber \\
&&s_{b}\,[B + \bar B + (C\times \bar C) ] = i\,[B + \bar B + (C\times \bar C)]\times  C.
\end{eqnarray}
We note that $s_{(a)d}\,[{\cal B}\times C] = 0$ and $s_{(a)d}\,[{\cal B}\times \bar C] = 0$
(under the (anti-)co-BRST symmetry transformations) but this kind of {\it perfect} 
symmetry is {\it not} obeyed by the CF-condition
$B + \bar B + (C\times \bar C) = 0$ in the context of (anti-)BRST symmetries. The latter condition is 
(anti-)BRST invariant only on the hypersurface where the CF-condition is satisfied. We lay emphasis
on the fact that the (anti-)BRST symmetries are {\it true} for any arbitary non-Abelain 1-form
gauge theory {\it but} the (anti-)co-BRST symmetries exist {\it only} for the 2$D$ non-Abelain 1-form gauge theory.
Both these symmetries are physically interesting because both are used [21] to prove that the 2$D$ non-Abelian
1-form gauge theory (without any interaction with matter fields) is a {\it new} model of topological field theory
(TFT) which captures a few key properties of Witten-type TFTs and some salient features of Schwarz-type TFTs.

\section{Nilpotent (Anti-)BRST Transformations: Horizontality Condition }

We very concisely mention here the salient features of the horizontality condition (HC) that leads to the derivation 
of proper (anti-)BRST symmetry transformations as well as CF-condition ($B+\bar B+(C\times \bar C)=0$)
within the framework of {\it usual} superfield approach to BRST formalism. In this context, it is worthwhile to mention
that the geometrical strength of the curvature 2-form 
($F^{(2)}=\,dA^{(1)}+ i\,A^{(1)}\wedge A^{(1)}=\frac{(dx^\mu\wedge dx^\nu)}{2!} F_{\mu\nu}$) 
plays a crucial role in this technique where $F^{(2)}$ is generalized to the the suitably chosen
supermanifold as:
\begin{eqnarray}
F^{(2)}(x)\to\tilde F^{(2)}(x,\theta,\bar\theta) = \Bigl(\frac{dZ^M\,\wedge dZ^N}{2!}\Bigr)\,\tilde F_{MN}(x,\theta,\bar\theta).
\end{eqnarray}
In the above, $\tilde F^{(2)}$ is
the supercurvature 2-form which is defined on a ($D, 2$)-dimensional supermanifold corresponding to a given
$D$-dimensional non-Abelian 1-form gauge theory where 
$\tilde F^{(2)} = \tilde d \tilde A^{(1)} + i\,(\tilde A^{(1)}\wedge \tilde A^{(1)})$ and $Z^M = (x^\mu,\theta,\bar\theta)$
are the superspace coordinates.
In this expression, we have generalization of the exterior derivative $d=dx^\mu\partial_\mu$ and
1-form connection $A^{(1)}=dx^{\mu} A_\mu$ (defined on the $D$-dimensional flat Minkowski space) to ($D, 2$)-dimensional
supermanifold (on which the given $D$-dimensional gauge theory is generalized). In other words, we have
the following: 
\begin{eqnarray}
&&d\to \tilde d= dx^\mu\,\partial_\mu + d\theta\,\partial_\theta + d\bar\theta \,\partial_{\bar\theta}, 
\quad \qquad\quad \tilde d^2 = 0,\nonumber\\
&&A^{(1)}=dx^\mu\,A_\mu (x)\to \tilde A^{(1)} = 
dx^{\mu}\,B_{\mu}(x, \theta, \bar\theta) + d\theta\,\bar F(x, \theta, \bar\theta)
+ d\bar\theta\,F(x, \theta, \bar\theta),
\end{eqnarray}
where [$B_{\mu}(x, \theta, \bar\theta),F(x, \theta, \bar\theta),\bar F(x, \theta, \bar\theta)$] are the superfields
on the ($D,2$)-dimensional supermanifold corresponding to the ordinary fields 
[$A_{\mu}(x), C(x),\bar C(x)$] of the $D$-dimensional non-Abelian gauge
theory. The above supermanifolds have the following expansions along the 
Grassmannian directions of the ($D, 2$)-dimensional supermanifold [5-7], namely; 
\begin{eqnarray}
&&A_\mu(x)\to B_\mu (x,\theta,\bar\theta) = A_\mu (x) + \theta \,\bar R_\mu (x)
+ \bar \theta \, R_\mu (x) + i\, \theta \bar \theta \,S_\mu (x), \nonumber \\
&&C(x)\to F(x,\theta,\bar\theta) = C(x) + i\, \theta \,\bar B_1 + i\, \bar\theta \,B_1 + i\, \theta \bar \theta \,s(x), \nonumber \\
&&\bar C(x) \to \bar F(x,\theta,\bar\theta) = \bar C(x) + i \,\theta\, \bar B_2 +  i\, \bar \theta \, B_2 
+ i \,\theta \bar \theta \, \bar s(x),
\end{eqnarray}
where, on the r.h.s., we have ($S_\mu, B_1, \bar B_1, B_2, \bar B_2$) and ($R_\mu, \bar R_\mu, S, \bar S$) as
the secondary fields which are bosonic and fermionic in nature, respectively. These secondary fields are determined
in terms of the basic and auxiliary fields of the theory due to the beauty and strength of HC.
We elaborate below some of the key features of HC in a concise manner.

We observe that the kinetic term $(-\frac{1}{4}\,F_{\mu\nu}\cdot F^{\mu\nu})$ remains invariant
under the nilpotent (anti-)BRST symmetries (because, primarily, it is a gauge invariant quantity). Within the framework
of USFA to BRST formalism, we demand that all the Grassmannian components of the (anti-)symmetric
supercurvature tensor 
$\tilde F_{MN} = (\tilde F_{\mu\theta}, \tilde F_{\mu\bar\theta}, \tilde F_{\theta\theta}, 
\tilde F_{\theta \bar\theta}, \tilde F_{\bar\theta \bar\theta})$
should be set equal to zero so that we have the following equality of the kinetic term (cf. Eq. (11))
\begin{eqnarray}
-\frac{1}{4}\,\tilde F_{MN}(x,\theta,\bar\theta)\cdot \tilde F^{MN}(x,\theta,\bar\theta) 
= -\frac{1}{4}\,F_{\mu\nu}(x)\cdot F^{\mu\nu}(x),
\end{eqnarray}
which is a gauge invariant restriction (GIR).
To achieve the equality (14), one of the  simplest choices is to set all the Grassmanian components of 
$\tilde F_{MN}(x,\theta,\bar\theta)$ equal
to zero so that only the antisymmetric spacetime components $\tilde F_{\mu\nu}(x,\theta,\bar\theta)$ survive.
To be precise, the restrictions
$\tilde F_{\mu\theta}= \tilde F_{\mu\bar\theta} = \tilde F_{\theta\theta}= \tilde F_{\theta \bar\theta}= 
\tilde F_{\bar\theta \bar\theta} = 0$
lead to the following relationship between the secondary fields
and basic as well as auxiliary fields (with the identifications $\bar B_1=\bar B, B_2 = B$), namely;
\begin{eqnarray}
&&R_\mu = D_\mu C, \quad \bar R_\mu = D_\mu \bar C, \quad 
S_\mu  = (D_\mu B + D_\mu C\times \bar C)
\equiv  -(D_\mu \bar B + C\times D_\mu \bar C), \nonumber \\ 
&&s = i \,(\bar B\times C), \quad \quad
\bar s = -i\,(B\times \bar C), \quad \quad
B_1 = -\frac{1}{2}\,(C \times C),\nonumber \\
&&\bar B_2 = -\frac {1}{2}\,(\bar C\times \bar C), \quad \quad
B+ \bar B + (C\times \bar C) = 0,
\end{eqnarray}
where the {\it last} entry is nothing but the celebrated CF-condition [23]. Thus, it is the theoretical strength of 
HC that we have determined {\it all} the secondary fields in terms of the basic and auxiliary fields of the theory
described by the Lagrangian density (1).

The substitution of the expressions for the above secondary fields into the expansions (13) leads to the following
super expansions of the superfields [5-7]
\begin{eqnarray}
B_{\mu}^{(h)}(x, \theta, \bar\theta) &= &A_{\mu}(x) + \theta \, (D_{\mu}\bar C) + \bar\theta \,(D_\mu C) + 
i\,\theta\bar\theta \, [D_{\mu}B + D_\mu C\times \bar C] \\ \nonumber
&\equiv & A_{\mu}(x) + \theta \, (s_{ab}A_{\mu}) + \bar\theta \,(s_{b}A_{\mu}) + \theta\bar\theta \,(s_{b}s_{ab}A_{\mu}), 
\\ \nonumber
F^{(h)}(x, \theta, \bar\theta) &= & C(x) + \theta \,(i \bar B) + \bar\theta \, (-\frac{i}{2}(C\times C)) + 
\theta\bar\theta \, (- \bar B\times  C) \\ \nonumber
&\equiv & C(x) + \theta \, (s_{ab} C) + \bar\theta \,(s_{b} C) + \theta\bar\theta \,(s_{b}s_{ab} C), \\ \nonumber
\bar F^{(h)}(x, \theta, \bar\theta) &= & \bar C(x)  
+ \theta \,(-\frac{i}{2}(\bar C\times \bar C)) + \bar\theta \,(i B)
+ \theta\bar\theta \,(B\times \bar C) \\ \nonumber
&\equiv & \bar C(x) + \theta \,(s_{ab} \bar C) + \bar\theta \,(s_{b} \bar C) + \theta\bar\theta \,(s_{b}s_{ab}\bar C),
\end{eqnarray}  
where the superscript $(h)$ denotes that the above superfields have been determined after the
application of HC. We note that the coefficients of $\theta, \bar \theta$ and $\theta \bar\theta$ are
nothing but the (anti-)BRST symmetries (3) of the $D$-dimensional non-Abelian gauge theory (without any
interactions with matter fields). In other words, we observe that the HC leads to the 
determination of proper (i.e. off-shell
nilpotent and absolutely anticommuting) (anti-)BRST symmetry transformations for the (anti-)ghost and
gauge fields of the non-Abelain 1-form gauge theory in any {\it arbitrary} dimension of flat Minkowski spacetime.

We end this section with the following important remarks. First of all, it is clear 
that the kinetic term remains
invariant under the (anti-)BRST symmetry transformations. The curvature tensor 
$F_{\mu\nu} = \partial_{\mu}A_{\nu} - \partial_{\nu}A_{\mu} + i \,(A_{\mu}\times A_{\nu})$ owes its origin
to the exterior derivative ($d = dx^{\mu}\partial_{\mu}$) because it is derived from the curvature 2-form
$F^{(2)} = dA^{(1)} + i\, A^{(1)}\wedge A^{(1)}$. Second, we note that the CF-condition is responsible
for the existence of the coupled (but equivalent) Lagrangian densities (1) for the non-Abelian
1-form gauge theory. The equivalence can be checked from our observations in Eq. (4). Thus, it is 
evident that {\it both} the Lagrangian densities respect the (anti-)BRST symmetry transformations 
{\it only} on the hypersurface which is described in the language of CF-condition. Third, we also
observe that the absolute anticommutativity property of $s_{(a)b}$ (i.e. $s_b s_{ab} + s_{ab} s_b = 0$)
is satisfied if and only if we use the CF-condition (which is one of the hallmarks of a {\it quantum} gauge
theory when it is described within the framework of BRST formalism). Fourth, we note that the CF-condition
is an (anti-)BRST invariant (i.e. $s_{(a)b} \, [B + \bar B + (C\times \bar C)] = 0$) 
quantity at the {\it quantum} level (cf. Eq. (10)).
Hence, this condition is a {\it physical} restriction.
Fifth, we point out that the surviving component of the super curvature tensor is equal to:
\begin{eqnarray} 
\tilde F_{\mu\nu}^{(h)}(x, \theta, \bar\theta) & = &\partial_\mu\, B_\nu^{(h)} - \partial_\nu\, B_\mu^{(h)} 
+ i\,(B_\mu^{(h)}\times  B_\nu^{(h)}) 
\equiv  F_{\mu\nu}(x) + \theta\,(i\,F_{\mu\nu}\times \bar C) \nonumber \\
& + & \bar \theta\,(i\,F_{\mu\nu}\times  C) 
+ \theta\bar\theta \,[-(F_{\mu\nu}\times C)\times \bar C - F_{\mu\nu}\times B].
\end{eqnarray}
The above equation demonstrates that $s_b\, F_{\mu\nu} = i\,(F_{\mu\nu}\times C)$, 
$s_{ab}\,F_{\mu\nu} = i\,(F_{\mu\nu}\times \bar C)$ and
$s_b\,s_{ab}\,F_{\mu\nu} = -\,\Bigl[ (F_{\mu\nu}\times B ) + (F_{\mu\nu}\times C)\times \bar C \Bigr]
\equiv \Bigl[(F_{\mu\nu}\times \bar B) + (F_{\mu\nu}\times \bar C)\times C\Bigr]$. It is self-evident that, for 
the 2$D$ non-Abelian theory (where $F_{\mu\nu}$ has only one existing component $F_{01} = E$), 
we have the following (anti-)BRST transformations for the field 
$E(x) = F_{01} = \partial_0 A_1(x) - \partial_1 A_0(x) + i\,(A_0(x)\times A_1(x))$, namely
\begin{eqnarray} 
E(x)\to \tilde E^{(h)}(x,\theta, \bar\theta) & = & E(x) + \theta\,(i\,E\times \bar C) + \bar\theta\,(i\,E\times C) \nonumber \\
& + &\,\theta\bar\theta\,[-(E\times C)\times \bar C - E\times B],
\end{eqnarray}  
which implies that $s_b\,E= i\,(E\times C), \,s_{ab}\,E= i\,(E\times \bar C), \,
s_b\,s_{ab}\,E = -\,\Bigl[(E\times B )+ (E\times C)\times \bar C 
\equiv \bigl[(E\times \bar B) + (E\times \bar C)\times C\big]\Bigr]$.
As a side remark, we note that this expression would turn out to be useful, later on, in Sec. 5.  Sixth, rest of the (anti-)BRST symmetry
transformations in (3) are determined due to the requirements of nilpotency and anticommutativity 
(which are the key properties of (anti-)BRST symmetries).
Seventh, it is evident from Eq. (16) that the (anti-)BRST symmetry transformations $s_{(a)b}$ are intimately
connected with the translational generators ($\partial_\theta, \partial_{\bar\theta}$). Eighth, the
nilpotency ($\partial_{\theta}^2 = \partial_{\bar\theta}^2 = 0$) and absolute anticommutativity 
($\partial_\theta\,\partial_{\bar\theta} + \partial_{\bar\theta}\,\partial_\theta = 0$) of the translational
generators ($\partial_\theta, \partial_{\bar\theta}$) imply that $s_{(a)b}^2 = 0$ and $s_b\,s_{ab} + s_{ab}\,s_b = 0$, too.
Finally, we note that 
$-\frac{1}{4}\,\tilde F^{(h)}_{\mu\nu}(x,\theta,\bar\theta)\cdot \tilde F^{\mu\nu (h)}(x,\theta,\bar\theta) 
= -\frac{1}{4}\,F_{\mu\nu}(x)\cdot F^{\mu\nu}(x)$ as desired (right from the beginning).

\section{(Anti-)co-BRST Symmetries: Superfield Formalism}

We derive here the (anti-)co-BRST symmetry transformations (6) by exploiting the ideas of AVSA. In this
connection, first of all, we take the generalization of the exterior derivative $d = dx^\mu \partial_\mu$ and ordinary 1-form
connection $A^{(1)} = dx^\mu \,(A_\mu.T)$ onto the (2, 2)-dimensional supermanifold as given in (12). 
We note that the gauge-fixing term of Lagrangian density (1) owes its origin to the co-exterior
derivative $(\delta)$, namely;
\begin{eqnarray}
\delta A^{(1)} = -*d * A^{(1)} = \partial_\mu A^\mu,\qquad \quad \delta  = - * \,d \,*, 
\end{eqnarray}
where $*$ is the Hodge duality operation on the 2{\it D ordinary} flat Minkowski spacetime manifold.
One of the salient features of the (anti-)co-BRST symmetry transformations (6) is the observation that it is
the gauge-fixing term (owing its origin to the co-exterior
derivative $\delta  = - * d *$) that remains invariant under them. The relation (19) can be generalized
onto our chosen (2, 2)-dimensional supermanifold. Thus, we invoke the following dual-horizontality condition (DHC)
(i.e. an analogue of HC), namely;
\begin{eqnarray}
\tilde \delta \tilde A^{(1)} = \delta A^{(1)}, \qquad \quad
\tilde\delta \tilde A^{(1)} = - \star \tilde d \star \tilde A^{(1)}, 
\end{eqnarray}
where $\tilde\delta = -\,\star\,\tilde d \,\star$ is the generalization of the ordinary co-exterior derivative 
$\delta  = - *\, d \,* $ onto the above chosen supermanifold and $\star$ is the Hodge duality operation on the (2, 2)-dimensional supermanifold. For the Abelian 1-form theory, the $\star$ operator has been defined explicitly in [24] 
on the (2, 2)-dimensional supermanifold.

In our Appendix A, the step-by-step computation of the l.h.s. of the DHC ($\tilde \delta \tilde A^{(1)} = \delta A^{(1)}$)
has been worked out. We take that result and write it in the following fashion:
\begin{eqnarray}
[\partial_\mu B^\mu + \partial_\theta \bar F + \partial_{\bar\theta} F ] + 
s^{\bar\theta\bar\theta}\,(\partial_{\bar\theta} \bar F) + s^{\theta\theta} \,(\partial_\theta F) = \partial_\mu A^{\mu}.
\end{eqnarray}
The above equality yields the relationships as listed below:
\begin{eqnarray}
\partial_\theta \, F = 0,\,\qquad \quad \partial_{\bar\theta} \,\bar F = 0,\,\qquad \quad
\partial_\mu B^{\mu} + \partial_\theta \bar F + \partial_{\bar \theta} F = \partial_\mu A^{\mu}.
\end{eqnarray}
This is due to the fact that there is {\it no} presence of factors like $s^{\theta\theta}$ and $s^{\bar\theta\bar\theta}$
on the r.h.s.
At this stage, we have to take into account the expansion of $B_\mu (x,\theta, \bar\theta)$, $F(x,\theta,\bar\theta)$
and $\bar F(x,\theta,\bar\theta)$ along the Grassmannian directions $(\theta,\bar\theta)$ of the (2, 2)
dimensional supermanifold as given in (13). 
Their substitution leads to the following restrictions from (22), namely; 
\begin{eqnarray}
&& \partial_\mu \bar R^{\mu} = 0,\,\qquad  \partial_\mu R^{\mu} = 0,\,\qquad  \partial_\mu S^{\mu} = 0,
\quad s = 0, \,\nonumber\\
&& \bar B_1 = 0,  \,\qquad B_2 = 0,\,\qquad \bar s = 0,\qquad B_1 + \bar B_2 = 0, 
\end{eqnarray}
where $ 
B_1 + \bar B_2 = 0 $ is the analogue of the CF-type condition. We make the choice $B_1 = -{\cal B}$ which
implies that $\bar B_2 = {\cal B}$. Thus, we obtain the expansions of the fermionic superfields 
$F(x,\theta,\bar\theta)$ and  ${\bar F}(x,\theta,\bar\theta)$
along the Grassmannian directions $(\theta, \bar\theta)$ as follows
\begin{eqnarray}
&&F^{(d)}(x,\theta,\bar\theta) = C(x) + \bar \theta \,(-i\,{\cal B}) \equiv C(x) + \bar \theta\,(s_d C), \nonumber \\
&&\bar F^{(d)}(x,\theta,\bar\theta) = \bar C(x) + \theta \,(i\,{\cal B}) \equiv \bar C(x) + \theta \,(s_{ad} \bar C),
\end{eqnarray}
where the superscript $(d)$ stands for the expansions of the superfields, obtained after the application 
of the DHC, and $s_{(a)d}$ are the (anti-)co-BRST symmetry transformations for the fields $(\bar C) C$ 
that have been quoted in Eq. (6). We note, at this juncture, that we have already derived the (anti-)co-BRST symmetry
transformations for the (anti-)ghost fields $(\bar C) C$ of our theory (cf. Eq. (6)). It is also
clear that $\partial_\theta \bar F^{(d)} = s_{ad}\bar C$ and $\partial_{\bar\theta} F^{(d)} = s_{d} C$.
These relationships show that the (anti-)co-BRST symmetry transformations $s_{(a)d}$ can be identified
with the translational generators ($\partial_\theta, \partial_{\bar\theta} $) along the Grassmannian 
directions ($\theta, \bar\theta$), respectively. Moreover, the above identifications imply that $s^2_{(a)d} = 0$ due to 
$\partial^2_\theta = \partial^2_{\bar\theta} = 0$ and $s_d\,s_{ad} + s_{ad}\,s_d = 0$ because of
$\partial_\theta\,\partial_{\bar\theta} + \partial_{\bar\theta}\,\partial_\theta = 0$.

We have to compute now the (anti-)co-BRST symmetry transformations for the gauge field
$A_\mu \equiv (A_\mu \cdot T)$. For this purpose, we have to exploit the ideas behind the AVSA
where the (anti-)co-BRST invariant quantities would be required to be independent of the ``soul'' coordinates
($\theta, \bar\theta$). In this context, we observe that the following
intresting and useful combination of fields
(in the square bracket below) is an (anti-)co-BRST invariant quantity, namely;
\begin{eqnarray}  
s_{(a)d}\,\Bigl [\varepsilon^{\mu\nu} A_\nu\cdot \partial_\mu{\cal B} - i\,\partial_\mu\bar C\cdot \partial^\mu C \Bigr] = 0.
\end{eqnarray} 
According to the basic tenets of AVSA, we have to equate the quantity in the square bracket with its
counterparts in terms of the superfields, namely;  
\begin{eqnarray}  
\varepsilon^{\mu\nu} B_\nu(x, \theta,\bar\theta)\cdot \partial_\mu{\cal B}(x) 
- i\,\partial_\mu \bar F^{(d)}(x,\theta,\bar\theta)\cdot \partial^{\mu}F^{(d)}(x,\theta,\bar\theta) \nonumber \\
= \varepsilon^{\mu\nu}A_\nu(x)\cdot\partial_\mu{\cal B}(x) - i\,\partial_\mu\bar C(x)\cdot \partial^\mu C(x),
\end{eqnarray}  
where we have taken $F^{(d)}$ and $\bar F^{(d)}$ from Eq. (24) and 
 ${\cal B}(x)\to {\cal B}(x,\theta,\bar\theta) = {\cal B}(x)$  because of the fact that
$s_{(a)d}\,{\cal B}(x) = 0$. Hence, it will have {\it no} expansion along 
$(\theta, \bar\theta)$-directions of the (2, 2)-dimensional supermanifold
provided we accept the result that the coefficients of $\theta$ and $\bar\theta$ correspond
to the (anti-)co-BRST symmetry transformations, respectively. The above restriction
(cf. Eq. (26)) is called as the dual-gauge invariant restriction (DGIR). Physically, this restriction
implies that the (anti-)co-BRST invariant quantities should remain
independent of the ``soul'' coordinates ($\theta, \bar\theta$) because the latter are
{\it only} the mathematical artifacts. It will be noted that
we have taken the expansions from (24) for the generalization of the fields $C(x)$ and $\bar C(x)$.
The explicit substitutions, from (24) into (26), yield: 
\begin{eqnarray}
\varepsilon^{\mu\nu}\bar R_\nu + \partial^\mu C = 0, \qquad \varepsilon^{\mu\nu} R_\nu + \partial^\mu \bar C = 0, \qquad 
\varepsilon^{\mu\nu} S_\nu - \partial^\mu {\cal B} = 0.
\end{eqnarray}  
The above relations lead to the explicit and exact derivation of the secondary fields of the expansions (that are present for
$B_\mu(x,\theta,\bar\theta)$ in Eq. (13)) as: 
\begin{eqnarray} 
R_\mu = -\,\varepsilon_{\mu\nu}\partial^\nu\bar C,\qquad \bar R_\mu = -\,\varepsilon_{\mu\nu}\partial^\nu C,\qquad
S_\mu = \varepsilon_{\mu\nu}\partial^\nu {\cal B}.
\end{eqnarray}  
Thus, ultimately, we obtain the expansions of the superfield $B_\mu(x,\theta,\bar\theta)$  as
\begin{eqnarray}
B_\mu^{(dg)}(x,\theta,\bar\theta) & = & A_\mu(x) + \theta \,(-\varepsilon_{\mu\nu}\partial^\nu C)
+ \bar\theta\, (-\varepsilon_{\mu\nu}\partial^\nu \bar C) 
+ \theta\bar \theta\,\, (i\,\varepsilon_{\mu\nu}\partial^\nu {\cal B}) \nonumber \\
& \equiv & A_\mu(x) + \theta \,(s_{ad} A_\mu )
+ \bar\theta\, (s_d  A_\mu ) 
+ \theta\bar \theta\,\, (s_d s_{ad} A_\mu ),
\end{eqnarray}  
where the superscript $(dg)$ denotes that the above expansion has been obtained
after the application of the dual-gauge invariant restriction
(DGIR). The expansion in Eq. (29) leads to the derivations of (anti-)co-BRST symmetry transformations for the
gauge field $A_\mu \equiv (A_\mu\cdot T)$ (cf. Eq. (6)) as the coefficients of the
Grassmanian variables $\theta$ and $\bar\theta $. The noteworthy point is the fact
that DHC and DGIR are intertwined {\it together} in a very useful fashion in Eq. (26). This is
the beauty and strength of AVSA.

We note, from Eq. (6), that the component 
$F_{01} = E =  -\,\varepsilon^{\mu\nu}\,\bigl(\partial_\mu\,A_\nu + \frac{i}{2}\,A_\mu\times A_\nu \bigr)$ of the curvature tensor
$F_{\mu\nu} = \partial_\mu\,A_\nu - \partial_\nu\,A_\mu + i\,(A_\mu\times A_\nu)$ transforms
under the (anti-)co-BRST symmetry transformations as: $s_d\,E = D_\mu\,\partial^\mu\,\bar C$ and 
$s_{ad} E = D_\mu\,\partial^\mu C$. These transformations can be derived from the AVSA to BRST formalism as:
\begin{eqnarray}
E(x)\to\tilde E(x, \theta,\bar\theta) & = &- \,\varepsilon^{\mu\nu}\,
\Bigl[ \partial_\mu\,B_\nu^{(dg)} + \frac{i}{2}\,(B_\mu^{(dg)}\times B_\nu^{(dg)})\Bigr] \nonumber \\
&\equiv &E(x) + \theta\,(D_\mu\,\partial^\mu C) + \bar\theta \,(D_\mu\,\partial^\mu \bar C) \nonumber \\
&+ &\theta \bar\theta \Bigl[-\,i\,D_\mu\,\partial^\mu {\cal B} - i\,\varepsilon_{\mu\nu}\,\partial^\nu \bar C\times \partial^\mu C\Bigr].
\end{eqnarray}  
We observe that the coefficients of $\theta, \bar \theta$ and $\theta\bar\theta$ {\it do} lead
to the derivation of $s_{ad} E(x),\, s_d\,E(x) $  and $s_ds_{ad} E(x)$. In other words, we obtain
$\partial_\theta \tilde E (x, \theta, \bar\theta)| _{\bar \theta=0} = s_{ad}\, E(x)$ and
$\partial_{\bar\theta} \tilde E (x, \theta, \bar\theta)| _{\theta=0} = s_{d}\, E(x)$ and
$\partial_{\bar\theta}\,\partial_\theta\,\tilde E(x, \theta, \bar\theta) = s_d\,s_{ad} \,E(x)$
which imply that the translational generators ($\partial_\theta, \partial_{\bar\theta}$)
correspond to the nilpotent (anti-)co-BRST symmetry transformations ($s_{a)d}$). Thus, we note that
$\partial_\theta^2 = \partial_{\bar\theta}^2 = 0$ and 
$\partial_\theta\,\partial_{\bar\theta} + \partial_{\bar\theta}\,\partial_\theta = 0$ are 
intimately connected with the nilpotency
(i.e. $s_{(a)d}^2=0$) and absolute anticommutativity (i.e. $s_d\,s_{ad} + s_{ad}\,s_d = 0$) properties
of the (anti-)co-BRST symmetry transformations ($s_{(a)d}$).

\section{CF-Type Restrictions: Superfield Approach}  
  
We discuss here the derivation of {\it all} possible CF-type restrictions that could be, in general, 
supported by the 2$D$ non-Abelian theory
within the framework of AVSA. First of all, we observe {\it here} that these physically motivated restrictions
have been derived in our
earlier work [22] by exploiting the idea of (anti-)BRST and (anti-)co-BRST symmetry invariance. Thus, our central goal is to derive them
by demanding, first of all, that the original CF-condition [$B + \bar B + (C\times \bar C) = 0 $] should be invariant under
the (anti-)co-BRST symmetry transformations. In the language of AVSA, we demand that this condition
should be valid on the (2, 2)-dimensional supermanifold, too, namely;
\begin{eqnarray}
B(x) + \bar B(x) + [F^{(d)}(x, \theta, \bar\theta)\times \bar F^{(d)}(x, \theta, \bar\theta)] = 
B(x) + \bar B(x) + [C(x)\times \bar C(x)], 
\end{eqnarray}  
where we have taken the generalizations:  $B(x)\to \tilde B(x, \theta, \bar\theta) = B(x), \quad
\bar B(x)\to \tilde {\bar B}(x, \theta, \bar\theta) = \bar B(x)$ due to the fact
that {\it both} these auxiliary fields are (anti-)co-BRST invariant quantities 
(i.e. $s_{(a)d} B(x) = 0,\,\,s_{(a)d} \bar B(x) = 0 $). In other words, the superfields $\tilde B(x,\theta,\bar\theta)$
and $\tilde{\bar B}(x,\theta,\bar\theta)$ have {\it no} expansions along $\theta$ and $\bar\theta$ directions of the
(2, 2)-dimensional supermanifold. Plugging in the expansions from
Eq. (24), we obtain (from the above) the following 
\begin{eqnarray}
i\,\theta \,({\cal B}\times C) - i\,\bar\theta \,({\cal B}\times \bar C) - \theta\,\bar \theta\,({\cal B}\times {\cal B}) = 0,
\end{eqnarray} 
which leads to the CF-type restrictions:  ${\cal B}\times  C = 0$ and ${\cal B}\times \bar C = 0$ (because  
${\cal B}\times {\cal B} = 0$ automatically). We observe that these
CF-type restrictions have appeared earlier in Eqs. (8) and (9) in the Lagrangian formulation.
We further observe, at this stage, that these {\it new} CF-type restrictions
are invariant under the nilpotent (anti-)co-BRST symmetry transformations $s_{(a)d}$ [i.e. $s_{(a)d}\,({\cal B}\times C) = 0,\,\,
s_{(a)d}\,({\cal B}\times \bar C) = 0$]. However, these are {\it not} invariant under the (anti-)BRST symmetry 
transformations. To have (anti-)BRST and (anti-)co-BRST symmetries {\it together}
in the 2$D$ theory, we have to 
demand that these {\it new} CF-type restrictions (${\cal B}\times C = 0,\,\,{\cal B}\times \bar C = 0$) should {\it also}
remain invariant under the (anti-)BRST symmetry transformations. Thus, within the framework of 
AVSA to BRST formalism, we demand the following equalities, namely;
\begin{eqnarray}
&&\tilde {\cal B}^{(g)}(x, \theta, \bar\theta)\,\times F^{(h)}(x, \theta, \bar\theta) = {\cal B}(x) \,\times C(x),\nonumber \\
&&\tilde {\cal B}^{(g)}(x, \theta, \bar\theta)\,\times \bar F^{(h)}(x, \theta, \bar\theta) = {\cal B}(x) \,\times \bar C(x),
\end{eqnarray}   
where the expansions for $F^{(h)}(x, \theta, \bar\theta)$ and $\bar F^{(h)}(x, \theta, \bar\theta)$ are
given in Eq. (16) which have been obtained after the application of HC. The expansions for 
$\tilde{\cal B}^{(g)}(x, \theta, \bar\theta)$ along the Grassmannian directions ($\theta, \bar\theta$) can be written as 
\begin{eqnarray}
\tilde{\cal B}^{(g)}(x, \theta, \bar\theta) = {\cal B} + \theta\,[i\,({\cal B}\times \bar C] 
+ \bar\theta\,[i\,({\cal B}\times  C] + \theta\,\bar\theta\,
\Bigl[-\,({\cal B}\times  B) - ({\cal B}\times  C)\times \bar C \Bigr],
\end{eqnarray}   
in view of the (anti-)BRST symmetry transformations: 
$s_b\,{\cal B} = i\,({\cal B}\times  C),\,\,s_{ab}\,{\cal B} = i\,({\cal B}\times \bar C) $ and
$s_b\,s_{ab}\,{\cal B} = [-\,({\cal B}\times  B) - ({\cal B}\times  C)\times \bar C ]$ on the auxilary
field ${\cal B}(x)$. The superscript $(g)$ on the superfield ${\cal B}(x,\theta, \bar\theta)$ has been 
taken into account to denote that the above expansion, for this superfield, has been derived due to the GIR which
we explain below.

We exploit the theoretical strength of AVSA to BRST formalism to determine the super expansion for the 
superfield ${\cal B}(x,\theta,\bar\theta)$. In this connection, we note that $s_{(a)b}\,[{\cal B}\cdot E] = 0$. Thus,
the basic tenets of AVSA permits us to demand the following equality
\begin{eqnarray}
{\cal B}(x, \theta, \bar\theta)\cdot\tilde E^{(h)}(x, \theta, \bar\theta ) = {\cal B}(x)\cdot E(x),
\end{eqnarray}
where the explicit expansion of $\tilde E^{(h)}(x,\theta,\bar\theta)$ has been quoted in Eq. (18). The substitution
of this result into the above equation implies that we have the following 
\begin{eqnarray}
&&P(x) = i\,({\cal B}\times \bar C), \qquad \qquad Q(x) = i\,({\cal B}\times C), \nonumber \\
&&M(x) = i\,\Bigl[({\cal B}\times B) + ({\cal B}\times C)\times \bar C \Bigr] 
\equiv  -\, i\,\Bigl[{\cal B}\times \bar B + ({\cal B}\times \bar C)\times C \Bigr],
\end{eqnarray}
where the fields $P(x), Q(x)$ and $M(x)$, in the above, are the secondary fields in the general super expansions of 
${\cal B}(x, \theta,\bar\theta)$
as given below:
\begin{eqnarray}
{\cal B}(x)\to\tilde{\cal B}(x,\theta,\bar\theta) = {\cal B}(x) + \theta\,P(x) +\bar \theta \, Q(x) + i\,\theta\bar\theta\, M(x).
\end{eqnarray} 
The results in (36) show that $P(x)$ and $Q(x)$ are fermionic in nature in contrast to the bosonic nature of 
$M(x)$. This observation is {\it also} consistent with the fermionic nature of the 
Grassmannian variables $(\theta, \bar\theta)$ that are present on the r.h.s. of the expansion (37).
Thus, it is crystal clear that we have obtained the (anti-)BRST symmetry transformations for the
auxiliary field ${\cal B}(x)$ from the GIR that has been quoted in (35).

We focus now on the explicit form of the restrictions (33). These can be expanded as:  
\begin{eqnarray}
&&\Bigl({\cal B}+ \theta\,\bigl[i\,({\cal B}\times \bar C) \bigr] 
+ \bar\theta\,[i\,({\cal B}\times  C)] + \theta\,\bar\theta \,
\Bigl[-\,({\cal B}\times  B) - ({\cal B}\times  C)\times \bar C \Bigr]\Bigr)\times \nonumber \\
&&\Bigl( C + \theta\,(i\,\bar B) + \bar \theta \,\bigl[\frac {-i}{2}\,( C\times C)\bigr ] 
+ \theta\,\bar\theta \,\bigl[-\bar B\times C \bigr ]\Bigr) = {\cal B}(x)\times C(x),
\end{eqnarray}   
\begin{eqnarray}
&&\Bigl({\cal B} + \theta\,[i\,({\cal B}\times \bar C] 
+ \bar\theta\,[i\,({\cal B}\times  C)] + \theta\,\bar\theta\, 
\Bigl[-\,({\cal B}\times  B) - ({\cal B}\times  C)\times \bar C \Bigr]\Bigr)\times \nonumber \\
&&\Bigl(\bar C + \theta\,\Bigl[\frac {-i}{2}(\bar C\times \bar C) \Bigr ] + \bar \theta \,[i\, B] 
+ \theta\,\bar\theta \,\bigl[ B\times C \bigr ] \Bigr) = {\cal B}(x)\times \bar C(x).
\end{eqnarray} 
Setting the coefficients of $\theta, \bar\theta $  and $\theta \bar\theta $ equal to zero (in
the above), we obtain the following restrictions
${\cal B}\times \bar B = 0,\,\, {\cal B}\times  B = 0$  where we have used 
${\cal B}\times  C = 0,\,\,{\cal B}\times \bar C = 0 $ and $B + \bar B + (C\times \bar C) = 0$
which are the {\it original} restrictions on the theory. We have done it because, 
to obtain the {\it new} CF-type restrictions, we have utilized already derived
earlier CF-type restrictions of our theory. 
We would like to point out that these restrictions have already
appeared earlier in the Lagrangian formulation [22]. To be more explicit, it can be  checked that the coefficient
of $\theta$ in (38) leads to the CF-type restrictions ${\cal B}\times \bar B = 0 $ provided we use
${\cal B}\times C = 0$ and $B + \bar B + (C\times \bar C) = 0$. We also mention here
that coefficient of $\bar\theta$ does not yield anything and the coefficient of $\theta\bar\theta$
produces ${\cal B}\times B = 0$ when we use the original restrictions ${\cal B}\times C = 0$
and $B + \bar B + (C \times \bar C) = 0$.

We note that the CF-type restriction ${\cal B}\times \bar B = 0$ and  ${\cal B}\times B = 0$ are,
once again, invariant under the (anti-)co-BRST symmetry transformations 
(i.e. $s_{(a)d}\,\Bigl[{\cal B}\times B \Bigr] = 0,\,\,s_{(a)d}\,\Bigl[{\cal B}\times \bar B \Bigr]= 0$). Thus, we demand their
invariance under the (anti-)BRST symmetries (in view of having {\it both} the (anti-)BRST and
(anti-)co-BRST symmetries together in the 2$D$ theory) with the following restrictions on the (super)fields:
\begin{eqnarray}
&&\tilde {\cal B}^{(g)}(x, \theta, \bar\theta)\,\times \tilde B^{(g)}(x, \theta, \bar\theta) 
= {\cal B}(x) \,\times B(x),\nonumber \\
&&\tilde {\cal B}^{(g)}(x, \theta, \bar\theta)\,\times \tilde {\bar B}^{(g)}(x, \theta, \bar\theta) 
= {\cal B}(x) \,\times \bar B(x),
\end{eqnarray}     
where, the generalizations and super expressions of ${\cal B}(x), B(x)$ and $\bar B(x)$ fields, onto the
(2, 2)-dimensional supermanifold are (34) and the following:
\begin{eqnarray} 
&&B(x) \to \tilde B^{(g)}(x, \theta, \bar\theta) 
= B(x) + \theta\,(i\,[B\times \bar C]) + \bar \theta \,(0) + \theta \,\bar\theta \, (0), \nonumber \\
&&\bar B(x) \to \tilde {\bar B}^{(g)}(x, \theta, \bar\theta) 
= \bar B(x) + \theta\,(0) + \bar \theta \,(i\,[\bar B\times  C]) + \theta \,\bar\theta \,(0).
\end{eqnarray}  
In the above, the superscript $(g)$ denotes the fact that the superfields $\tilde B^{(g)}(x,\theta, \bar\theta)$
and $\tilde {\bar B}^{(g)}(x,\theta, \bar\theta)$ corresponds to the superfields that could be obtained
after the application of GIRs. We elaborate the derivation of these
expansions (i.e. Eq. (41)) within the framework of AVSA. In this 
context, we note that $s_b\, B(x) = 0$ and $s_b\,(B\times \bar C)= 0$. Thus, we have the following GIRs
in the language of the quantities on the supermanifold
\begin{eqnarray}
\partial_{\bar\theta}\,\tilde B(x, \theta, \bar\theta) = 0, \quad\quad \partial_{\bar\theta}\,
\Bigl[\tilde B(x, \theta, \bar\theta)\times \bar F^{(h)}(x, \theta, \bar\theta)\Bigr] = 0,
\end{eqnarray}
where $\bar F^{(h)}(x, \theta,\bar\theta)$ has been expressed in Eq. (16) and the general super expansion of 
$\tilde B(x,\theta,\bar\theta)$ along the Grassmannian directions ($\theta, \bar\theta)$ is as follows:
\begin{eqnarray}
B(x)\to \tilde B(x,\theta,\bar\theta) = B(x) + \theta\,U(x) + \bar\theta\, V(x) + \theta\,\bar\theta\,S(x).
\end{eqnarray}
In the above, the pair ($U(x), V(x)$) are fermionic and $S(x)$ is a bosonic secondary field due to the fermionic
nature of $(\theta, \bar\theta)$ and bosonic nature of $B(x)$. We have also taken into account the mapping 
$\partial_{\bar\theta}\leftrightarrow s_b$ (cf. Sec. 3).
It will be noted that $\partial_{\bar\theta}\,\tilde B(x,\theta,\bar\theta) = 0$ implies that $V(x)= S(x) = 0$. Thus,
the reduced form of $\tilde B(x,\theta,\bar\theta)$ is $\tilde B^{(r)}(x,\theta,\bar\theta) = B(x) + \theta\,U(x)$.
Now the {\it second} restriction in (42) can be expressed as:
\begin{eqnarray}
&&\partial_{\bar\theta}\,\Bigl[\tilde B^{(r)}(x,\theta,\bar\theta)\,\times \, 
\bar F^{(h)}(x,\theta,\bar\theta)\Bigr] = \nonumber \\
&&\frac{\partial}{\partial \bar\theta}\,\Bigl[\{B(x) + \theta\,U(x)\}\times
\{\bar C + \theta\,\Bigl(\frac{-i}{2}\,(\bar C\times \bar C)\Bigr) + \bar \theta\,(iB)+\theta\,\bar\theta \,(B\times\bar C)\}\Bigr].
\end{eqnarray}
The above restriction produces the form of $\tilde B(x,\theta,\bar\theta)$ that has been quoted in Eq. (41)
as $\tilde B^{(g)}(x,\theta,\bar\theta)$. In an exactly similar fashion,
we observe that $s_{ab}\,\bar B = 0$ and $s_{ab}\,(\bar B\times C) = 0$. Thus, we have the following 
restrictions, within  the framework of AVSA, on supermanifolds:
\begin{eqnarray}
\partial_\theta\,\Bigl[\tilde {\bar B}(x, \theta, \bar\theta)\Bigr] = 0, \quad\quad \partial_\theta\,
\Bigl[\tilde {\bar B}(x, \theta, \bar\theta)\times  F^{(h)}(x, \theta, \bar\theta)\Bigr] = 0,
\end{eqnarray}
where the general expansion for the superfield $\tilde{\bar B}(x,\theta,\bar\theta)$ is 
\begin{eqnarray}
\bar B(x)\to \tilde{\bar B}(x,\theta,\bar\theta) = \bar B(x) + \theta \,K(x) + \bar\theta\, L(x) + i\,\theta\bar\theta\, N(x).
\end{eqnarray}
In the above, the pair ($K(x), L(x)$) are the fermionic secondary fields, $N(x)$ is bosonic
and we have taken into account $\partial_\theta \leftrightarrow s_{ab}$ (cf. Sec. 3). The first condition in (45)
leads to
\begin{eqnarray}
K(x) = 0 ,\qquad  N(x) = 0,\qquad \tilde {\bar B}(x,\theta,\bar\theta)\to 
\tilde {\bar B}^{(r)}(x,\theta,\bar\theta) = \bar B(x) + \bar\theta \,L(x),
\end{eqnarray}
where the superscript $(r)$ denotes the reduced form of $\tilde {\bar B}(x,\theta,\bar\theta)$. Now plugging in  the 
values of $F^{(h)}(x,\theta,\bar\theta)$ from Eq. (16) and $\tilde {\bar B}^{(r)}(x,\theta,\bar\theta)$,
we obtain (from (45)) the following 
\begin{eqnarray}
\frac{\partial}{\partial\theta}\,\Bigl[\Bigl(\bar B(x) + \bar\theta\,L(x)\Bigr)\times \Bigl(C(x) + \theta\,(i\,\bar B) + 
\bar\theta\,[-\frac {i}{2}\,(C\times C)] + \theta\bar\theta\,(-\bar B\times C)\Bigr)\Bigr] = 0,
\end{eqnarray}
which, ultimately, yields the value of $L(x) = i\,(\bar B\times C)$. Substitution of this value in 
$\bar B^{(r)}(x, \theta,\bar\theta)$ produces $\tilde{\bar B}^{(g)}(x, \theta,\bar\theta)$
which has been quoted in Eq. (41).
The above expansions agree with the (anti-)BRST symmetry transformations 
$s_b\,B = 0,\, s_{ab} \,B = i\,(B\times \bar C),\, s_{b} \,\bar B = i\,(\bar B\times C),\, s_{ab}\,\bar B = 0$  
which can also be derived by the requirements of the nilpotency and absolute anticommutativity properties . 
We lay emphasis on the fact that we have taken into account the super expansions of Sec. 3 and the 
identifications $s_b \leftrightarrow \partial_{\bar\theta}$ and $s_{ab} \leftrightarrow \partial_\theta$
for the derivation of Eq. (41).

The explicit substitutions, from (34) and (41) into (40), imply the following
equalities in the language of (super)fields on the supermanifold:  
\begin{eqnarray}
&&\Bigl({\cal B}+ \theta\,[i\,({\cal B}\times \bar C)] 
+ \bar\theta\,[i\,({\cal B}\times C] + \theta\,\bar\theta\,
\Bigl[-\,({\cal B}\times  B) - ({\cal B}\times  C)\times \bar C \Bigr]\Bigr)\times \nonumber \\
&&\Bigl({\cal B} + i\,\theta\,({\cal B}\times \bar C) \Bigr) 
 = {\cal B}(x)\times B(x),
\end{eqnarray} 
 
\begin{eqnarray}
&&\Bigl({\cal B} + \theta\,\bigl[i\,({\cal B}\times \bar C)\bigr] 
+ \bar\theta\,\bigl[i\,({\cal B}\times  C\bigr] + \theta\,\bar\theta\, 
\Bigl[-\,({\cal B}\times  B) - ({\cal B}\times  C)\times \bar C \Bigr]\Bigr)\times \nonumber \\
&&\Bigl(\bar B + i\,\bar \theta\,(\bar B\times \bar C)\Bigr) = {\cal B}(x)\times \bar B(x).
\end{eqnarray}   
The above equalities lead, ultimately, to the following {\it new} restrictions:  
\begin{eqnarray}
{\cal B}\times  B = 0,\,\qquad\bar B\times  C = 0, \,\qquad B\times \bar B = 0, \qquad B\times \bar C = 0.
\end{eqnarray}  
It is evident that the above restrictions are neither {\it perfectly} invariant under the (anti-) co-BRST
symmetries {\it nor} under the (anti-)BRST symmetry transformations. The substitutions of the superfields
$F^{(d)},\,\bar F^{(d)},\,F^{(h)},\,\bar F^{(h)},\,\tilde B^{(g)},\,\tilde {\bar B}^{(g)}, \,{\cal B}^{(g)}$ in a
straightforward manner (from appropriate equations) lead to the derivation of the new restrictions: 
\begin{eqnarray}
B\times C = 0, \,\quad\quad \bar B\times \bar C = 0.
\end{eqnarray} 
At this stage, the tower of restrictions terminate and there are {\it no} further CF-type restrictions on the 
2{\it D} non-Abelian 1-form gauge theory. Thus, we have derived here {\it all} possible CF-type restrictions that could 
be supported by the self-interacting 2{\it D} non-Abelian theory (without any interaction with matter fields).
It is pertinent to point out that the above tower of CF-type restrictions have been obtained in our
earlier work [22] on the basis of (anti-)BRST and (anti-)co-BRST symmetry invariance(s).

\section{Conclusions}

In our present endeavor, we have applied the geometrical AVSA to
BRST formalism for the derivation of proper (i.e. off-shell nilpotent
and absolutely anticomuting) (anti-)BRST and (anti-)co-BRST  transformations for the self-interacting
2$D$ non-Abelain 1-form gauge theory (without any interaction with matter fields). This exercise has been
specifically performed in the case of 2$D$ non-Abelian 1-form gauge theory 
where the (anti-)BRST and (anti-)co-BRST
symmetries co-exist {\it together} (see, e.g. [21]). In fact, these nilpotent symmetries prove that this 2$D$
non-Abelian model is a tractable physical example of the
Hodge theory as well as a {\it new} model of topological field theory (see, e.g. [21]). The latter claim is
true because it is corroborated by the observation that the 2{\it D} non-Abelian  theory
captures a few key properties of the Witten-type TFT and some salient
 features of the Schwarz-type TFT.
The decisive features of the above continuous symmetries (and their (anti)commutators) is the
observation that these {\it symmetries} (and their corresponding  {\it conserved charges}) provide the
physical realizations of the de Rham cohomological operators of differential geometry. Hence, our present
2$D$ self-interacting non-Abelian quantum field theoretic model turns out to be an example of the Hodge theory.

In our present investigation, we have applied the DHC and DGIR to obtain the (anti-) co-BRST symmetry
transformations for the 2$D$ non-Abelian 1-form gauge theory within the framework of AVSA to BRST 
formalism. This result is completely {\it novel} as, in our previous attempts [10-14], we have {\it not}
applied the AVSA to derive the (anti-)co-BRST symmetry transformations systematically
for the 2$D$ non-Abelian theory. Further, we 
have exploited the theoretical potential and power of the AVSA to BRST formalism to derive {\it all}
possible CF-type restrictions that could emerge from the {\it original} CF-condition
$(B+\bar B + (C\times \bar C) = 0)$ by demanding its invariance under the (anti-)BRST and (anti-)co-BRST 
symmetries. Of course, the latter requirements (i.e. (anti-)BRST and (anti-)co-BRST invariances) have been
expressed in the language of appropriately chosen superfields (that have been derived after the application
of (D)HCs and (D)GIRs). The upshot of this whole exercise is the emergence of a tower of CF-type restrictions that
could be, in general,  supported by the 2$D$ non-Abelain theory. It should be pointed out, however, that only 
{\it a few} of these CF-type restrictions have been {\it actually} utilized by us in the Lagrangian formulation
where the (anti-)BRST and (anti-)co-BRST symmetry transformations have been discussed.

We would like to comment on the possible existence of a tower of CF-type restrictions for our present 2$D$ non-Abelian
theory. First of all, we note that all these CF-type restrictions are in terms of the auxiliary fields and (anti-)ghost
fields. Thus, these restrictions do {\it not} affect the degrees of freedom
(d.o.f) counting of the gauge field ($A_\mu = A_\mu\cdot T$). We have
utilized only a few restrictions which include ${\cal B}\times C = 0$ and ${\cal B}\times \bar C = 0$ for the
(anti-) co-BRST invariance in our theory. It is pertinent to point out that $\{s_d, s_{ad}\} = 0$ implies that
co-BRST and anti-co-BRST symmetries are independent of 
each-other. Hence, the restrictions ${\cal B}\times C = 0$ and ${\cal B}\times \bar C = 0$ 
are independent of each-other and they do {\it not} imply that $C \times \bar C = 0$. Thus, the non-Abelian
nature of our theory remains intact 
(see, e.g. [22] for details).

In our earlier works (see, e.g. [15-17]), we have claimed that the (anti-)dual-BRST symmetry can exist
for the $p$-form ($p = 1, 2, 3,...$) gauge theories {\it only} in the 2$p$-dimensions of 
spacetime. Thus, for the (non-)Abelian 1-form gauge theories, the above (anti-)co-BRST symmetries exist 
{\it only} in 
two (1+1)-dimensions of spacetime. We have also demonstrated the existence of (anti-)co-BRST
symmetries in the cases of 4$D$ Abelian 2-form and 6$D$ Abelian 3-form gauge theories which 
corroborate the claims that have been made in our earlier work [16]. It would be a nice future
endeavor for us to apply the geometrical AVSA to BRST formalism for such theoretically interesting
systems to obtain the (anti-)co-BRST symmetries and tower of all CF-type restrictions. We would like to
add that we have also shown the existence of (anti-)co-BRST symmetries in the case of  a 1$D$ toy model
of a Hodge theory which is nothing but the model of a rigid rotor. It would be challenging to apply
the ideas of our present investigation to this 1$D$ system, too.
We are busy, at the moment, 
in exploring the proof of the above speculative ideas and our results would be reported  
elsewhere in our future publications [25].\\

\noindent
{\bf Competing Interests}\\

\noindent
The authors declare that they have no competing interests.\\

\noindent
{\bf Acknowledgements}\vskip 0.5cm
\noindent
N. Srinivas is grateful to the BHU-fellowship under which the present investigation has been carried out.   
Thanks  are also due to T. Bhanja for fruitful discussions.

\vskip 0.5cm 
\begin{center}
\Large{\bf Appendix A: On the Derivation of $\tilde \delta \tilde A^{(1)} = - \star \tilde d\,\star \tilde A^{(1)}$}\\
\end{center}  
We carry out here the step-by-step computation of the l.h.s. of the DHC (i.e. $\tilde \delta\,\tilde A^{(1)} = \delta A^{(1)}$)
by applying the Hodge duality $\star$ operation [24] on the (2, 2)-dimensional supermanifold that has been chosen for 
our discussions. First of all, we derive the following 3-form, namely;
\begin{eqnarray}
&&\star\,\tilde A^{(1)} = \star\,\Bigl[dx^\mu\,B_\mu + d\theta\,\bar F + d\bar \theta\,F \Bigr ] 
= \varepsilon^{\mu\nu}\,(dx_\nu\wedge d\theta\wedge d\bar\theta)\,B_\mu \nonumber \\
&&+ \frac{1}{2!}\,\varepsilon^{\mu\nu}\,(dx_\mu \wedge dx_\nu \wedge d\bar \theta)\,\bar F 
+ \frac{1}{2!}\,\varepsilon^{\mu\nu}\,(dx_\mu\wedge dx_\nu \wedge d\theta ) F,
\end{eqnarray}   
where we have used the following duality operations [24]  
\begin{eqnarray}
&&\star\,(dx^\mu)\,=\,\varepsilon^{\mu\nu}\, (dx_\nu\wedge d\theta\,\wedge\, d\bar\theta), \qquad
\star\,(d\theta)\,=\,\frac{1}{2!}\,\varepsilon^{\mu\nu}\, (dx_\mu\,\wedge dx_\nu\, \wedge\, d\bar\theta), \nonumber \\
&&\star\,(d\bar\theta)\,=\,\frac{1}{2!}\,\varepsilon^{\mu\nu}\, (dx_\mu\,\wedge dx_\nu\, \wedge\, d\theta),
\end{eqnarray}   
because of the fact that the Hodge dual of a 1-form, on a (2, 2)-dimensional supermanifold,
is a 3-form. Now we apply the super exterior derivative $\tilde d$ on (53) to obtain a 4-form on the
(2, 2)-dimensional supermanifold as:  
\begin{eqnarray}
\tilde d \,\star\,[\tilde A^{(1)}]& = &\varepsilon^{\mu\nu}\, 
(dx_\lambda\,\wedge\,dx_\nu\wedge d\theta\,\wedge\, d\bar\theta)\,\partial^\lambda B_\mu 
+\frac{1}{2!}\,\varepsilon^{\mu\nu}\,
(dx_\lambda\,\wedge\,dx_\mu\wedge\,dx_\nu\wedge\, d\bar\theta)\,\partial^\lambda \bar F  \nonumber \\
&+&\frac{1}{2!}\,\varepsilon^{\mu\nu}\, 
(dx_\lambda\,\wedge\,dx_\mu\wedge dx_\nu\,\wedge\, d\theta)\,\partial^{\lambda}  F 
-\,\varepsilon^{\mu\nu}\,(dx_\nu\,\wedge\,d\theta\,\wedge\, d\theta\,\wedge\, d\bar\theta)\,\partial_\theta B_\mu \nonumber \\
&-&\frac{1}{2!}\,\varepsilon^{\mu\nu}\, 
(dx_\mu\,\wedge\,dx_\nu\wedge d\theta\,\wedge\, d\bar\theta)\,\partial_\theta \bar F 
-\frac{1}{2!}\,\varepsilon^{\mu\nu}\, 
(dx_\mu\,\wedge\,dx_\nu\wedge d\theta\,\wedge\, d \theta)\,\partial_\theta  F \nonumber \\
&-&\,\varepsilon^{\mu\nu}\, 
(dx_\nu\wedge\,d\bar\theta \wedge\, d\theta\,\wedge\, d\bar\theta)\,\partial_{\bar\theta}  B_\mu 
-\frac{1}{2!}\,\varepsilon^{\mu\nu}\, 
(dx_\mu\,\wedge\,dx_\nu\wedge d\bar\theta\,\wedge\, d\bar\theta)\,\partial_{\bar\theta} \bar F \nonumber \\
&-&\frac{1}{2!}\,\varepsilon^{\mu\nu}\, 
(dx_\mu\,\wedge\,dx_\nu\wedge d\theta\,\wedge\, d\bar\theta)\,\partial_{\bar\theta} F.
\end{eqnarray}  
In the above, we have used the explicit expression of $\tilde d = dx^\mu\,\partial_\mu + d\theta\,\partial_\theta +d\bar\theta\,\partial_{\bar\theta}$  and have taken into account the anticommutativity property of the Grassmannian variables
($\theta, \bar\theta$) with their derivatives ($\partial_\theta, \partial_{\bar\theta}$) and 
the same property among themselves.

We are in the position now to apply an star ($-\,\star$) on the above 4-form to get a scalar (i.e. 0-form). Before we 
apply it, we would like to state that the {\it second} and {\it third} terms of (55) would be equal to 
{\it zero} because a 3-form
in spacetime differentials (i.e. $dx_\lambda\,\wedge\, dx_\mu\,\wedge\,dx_\nu$) can {\it not} exist on a (2, 2)-dimensional
supermanifold. Further, as per the rules of the Hodge duality $\star$ operation laid down in our earlier work [24],
we can {\it not} have the existence of 3-form differentials (e.g. $d\theta \wedge\,d\theta\,\wedge\,d\theta,\quad
d\bar\theta \wedge\,d\bar\theta\,\wedge\,d\bar\theta, \quad d\theta \wedge\,d\bar\theta\,\wedge\,d\bar\theta,\quad
d\theta \wedge\,d\theta\,\wedge\,d\bar\theta$) on the (2, 2)-dimensional supermanifold as it can accommodate {\it only}
2-form differentials in the Grassmannian variables (e.g. 
$d\theta\,\wedge d\theta,\,\,d\bar\theta\,\wedge d\bar\theta,\,\,d\theta\,\wedge d\bar\theta$). As a consequence, the 
{\it fourth} and {\it seventh} terms would be zero. Thus, the existing terms are:
\begin{eqnarray}
\tilde d \,\star\,[\tilde A^{(1)}]& = &\varepsilon^{\mu\nu}\, 
(dx_\lambda\,\wedge\,dx_\nu\wedge d\theta\,\wedge\, d\bar\theta)\,\partial^\lambda B_\mu 
-\frac{1}{2!}\,\varepsilon^{\mu\nu}\,
(dx_\mu\wedge\,dx_\nu\wedge\,d\theta \wedge d\bar\theta)\,\partial_\theta \bar F  \nonumber \\
&-&\frac{1}{2!}\,\varepsilon^{\mu\nu}\, 
(dx_\mu\,\wedge\,dx_\nu\wedge d\theta\,\wedge\, d\theta)\,\partial_\theta  F 
 -\frac{1}{2!}\,\varepsilon^{\mu\nu}\, 
(dx_\mu\,\wedge\,dx_\nu\wedge d\theta\,\wedge\, d\bar\theta)\,\partial_{\bar\theta}  F \nonumber \\
&-&\frac{1}{2!}\,\varepsilon^{\mu\nu}\, 
(dx_\mu\,\wedge\,dx_\nu\wedge d\bar\theta\,\wedge\, d\bar\theta)\,\partial_{\bar\theta}  \bar F. 
\end{eqnarray}  
Now, the application of ($-\star$) on the above equation leads to the derivation of Eq. (21)
(i.e. $\tilde \delta \tilde A^{(1)} = - \star \tilde d\,\star \tilde A^{(1)}$) that has been 
incorporated in our text. In this derivation, the following inputs have been used (see, e.g. [24] for details)  
\begin{eqnarray}  
&&\star\,(dx_\mu\,\wedge\,dx_\nu\wedge d\theta\,\wedge\, d\bar\theta) 
= \varepsilon_{\mu\nu}, \nonumber \\
&&\star\,(dx_\mu\,\wedge\,dx_\nu\wedge d\bar\theta\,\wedge\, d\bar\theta)
= \varepsilon_{\mu\nu}\, s^{\bar \theta\, \bar\theta}, \nonumber \\
&&\star\,(dx_\mu\,\wedge\,dx_\nu\wedge d\theta\,\wedge\, d\theta) 
= \varepsilon_{\mu\nu}\, s^{\theta\,\theta}, 
\end{eqnarray}   
where $s^{\theta\,\theta} $  and $s^{\bar\theta\,\bar\theta} $ are the factors that have been taken into account
so that another ($\star$) operation on (57) yields the original 4-forms with factor of $\pm$ signs
in front of them as per the rules of Hodge duality operator [24].


\begin{thebibliography}{99}
\bibitem{dirac}  See, e.g., C. N. Yang, {\it Physics Today} {\bf 33}, 42 (1980)
\bibitem{TM}     J. Thierry-Mieg, {\it J. Math. Phys.} {\bf 21}, 2834  (1980)
\bibitem{TM1}    J. Thierry-Mieg, {\it Nuovo Cimento} A {\bf 56}, 396 (1980)
\bibitem{QIR}    M. Quiros, F. J. de Urries, J. Hoyos, M. L. Mazon, E. Rodrigues,\\
                 {\it J. Math. Phys.} {\bf 22}, 1767 (1981)
\bibitem{BNR}    L. Bonora, M. Tonin, {\it Phys. Lett.} B {\bf 98}, 48 (1981)
\bibitem{BTP}    L. Bonora, P. Pasti, M. Tonin, {\it Nuovo Cimento} A {\bf 63}, 353 (1981)
\bibitem{BPT1}   L. Bonora,  P. Cotta-Ramusino, {\it Commun. Math. Phys.} {\bf 87}, 589 (1983)
\bibitem{DJT}    R. Delbourgo, P. D. Jarvis, G. Thompson, {\it Phys. Lett.} B {\bf 109}, 25 (1982)
\bibitem{LTM}    L. Baulieu, J. Thierry-Mieg, {\it Nucl. Phys.} B {\bf 197}, 477 (1982)
\bibitem{RPM2}   See, e.g., R. P. Malik, {\it Eur. Phys. J.} C {\bf 60}, 457 (2009)
\bibitem{RPM3}   See, e.g., R. P. Malik, {\it J. Phys. A: Math. Theor.} {\bf 39},  10575 (2006)
\bibitem{RPM4}   See, e.g., R. P. Malik, {\it Eur. Phys. J.} C {\bf 51}, 169 (2007)
\bibitem{RPM5}   See, e.g., R. P. Malik, {\it J. Phys. } A {\bf 37}, 5261 (2004)
\bibitem{RPM6}   See, e.g., R. P. Malik, {\it Eur. Phys. J.} C {\bf 48}, 825 (2006)
\bibitem{GM}     See, e.g., S. Gupta, R. P. Malik, {\it Eur. Phys. J.} C {\bf 58},  517 (2008)
\bibitem{KKM}    R. Kumar, S. Krishna, A. Shukla, R. P. Malik, \\
                 {\it Int. J. Mod. Phys.} A {\bf 29}, 1450135 (2014)
\bibitem{RPM12}  See, e.g., R. P. Malik, {\it Eur. Phys. J.} C {\bf 60}, 457 (2009)
\bibitem{KT2}    K. Nishijima, {\it Prog. Theor. Phys.} {\bf 80}, 905 (1988)
\bibitem{DVSP}   D. Dudal, V. E. R. Lemes, M. S. Sarandy, S. P. Sorella, M. Picariello,\\
                 {\it JHEP} {\bf 0212}, 008 (2002)
\bibitem{DVSP1}  D. Dudal, H. Verschelde, V. E. R. Lemes, M. S. Sarandy, S. P. Sorella, M. Picariello,
                 A. Vicini, J. A. Gracey, {\it JHEP} {\bf 0306}, 003 (2003)
\bibitem{RPM1}   See, e.g., R. P. Malik, {\it J. Phys. }A {\bf 34}, 4167 (2001)
\bibitem{NKM}    N. Srinivas, S. Kumar, B. K. Kureel, R. P. Malik, arXiv: 1606.05870 [hep-th]\\
(To appear in {\it  Int. J. Mod. Phys.} A (2017))
\bibitem{GF}     G. Curci, R. Ferrari, {\it Phys. Lett.} B {\bf 63}, 91 (1976)
\bibitem{RPM18}  See, e.g., R. P. Malik, {\it  Int. J. Mod. Phys.} A {\bf 21}, 3307 (2006)
\bibitem{RPM19}  R. P. Malik, {\it et al.,} in preperation
\end{thebibliography}
\end{document}